\documentclass[9pt, conference]{IEEEtran}
\IEEEoverridecommandlockouts

\usepackage{cite}
\usepackage{subcaption}
\usepackage{graphicx}
\usepackage{textcomp}
\usepackage{xcolor}
\usepackage{xspace}
\usepackage{amssymb}
\usepackage{hyperref}
\usepackage{cleveref}

\usepackage{booktabs}
\usepackage{tabularx}
\usepackage[labelformat=simple]{subcaption}
\usepackage{enumitem}
\usepackage[ruled, vlined,linesnumbered]{algorithm2e}
\usepackage{listings}
\usepackage{microtype}
\usepackage{xspace}
\usepackage[normalem]{ulem}

\SetKwProg{Function}{def}{:}{}

\newcommand{\work}{\textit{Tropical}\xspace}
\newcommand{\prefill}{\textit{prefill}\xspace}
\newcommand{\decode}{\textit{decode}\xspace}

\newcommand{\Prefill}{\textit{Prefill}\xspace}
\newcommand{\Decode}{\textit{Decode}\xspace}
\newcommand{\disaggregated}{\textit{disaggregated}\xspace}
\newcommand{\nondisaggregated}{\textit{non-disaggregated}\xspace}
\newcommand{\Disaggregated}{\textit{Disaggregated}\xspace}
\newcommand{\Nondisaggregated}{\textit{Non-disaggregated}\xspace}

\newcommand{\figref}[1]{Fig~\ref{#1}}
\newcommand{\secref}[1]{\sectionautorefname~\ref{#1}}

\DeclareRobustCommand{\IEEEauthorrefmark}[1]{\smash{\textsuperscript{\footnotesize #1}}}

\begin{document}

\title{\work: Enhancing SLO Attainment in Disaggregated LLM Serving via SLO-Aware Multiplexing {}
}
\author{
    \IEEEauthorblockN{
        Jinming Ma\IEEEauthorrefmark{1}\IEEEauthorrefmark{*},
        Jiefei Chen\IEEEauthorrefmark{1}\IEEEauthorrefmark{2}\IEEEauthorrefmark{*},
        Xiuhong Li\IEEEauthorrefmark{3}\IEEEauthorrefmark{\#},
        Jiangfei Duan\IEEEauthorrefmark{1}\IEEEauthorrefmark{4},
        Haojie Duanmu\IEEEauthorrefmark{1}\IEEEauthorrefmark{5}
    }
    \IEEEauthorblockN{
        Xingcheng Zhang\IEEEauthorrefmark{1}\IEEEauthorrefmark{6},
        Chao Yang\IEEEauthorrefmark{3},
        Dahua Lin\IEEEauthorrefmark{1}\IEEEauthorrefmark{4}
    }
    \IEEEauthorblockA{
        \IEEEauthorrefmark{1}Shanghai Artificial Intelligence Laboratory,
        \IEEEauthorrefmark{2}Fudan University,
        \IEEEauthorrefmark{3}Peking University
    }
    \IEEEauthorblockA{
    \IEEEauthorrefmark{4}The Chinese University of Hong Kong,
        \IEEEauthorrefmark{5}Shanghai Jiao Tong University,
        \IEEEauthorrefmark{6}Sensetime Research
    }
    \IEEEauthorblockA{
        Email: majinming@pjlab.org.cn, lixiuhong@pku.edu.cn
    }
    \thanks{
        \IEEEauthorrefmark{*} Equal contribution.
        \IEEEauthorrefmark{\#} Corresponding author.
    }
}
\maketitle
\begin{abstract}

To guarantee service quality in transformer based large language model (LLM) serving, it is essential to meet the latency constraints of both the \textit{prefill} phase (measured by Time-to-First-Token, TTFT) and the \textit{decode} phase (measured by Time-per-Output-Token, TPOT).
\Nondisaggregated serving places \prefill and \decode on the same worker, while \disaggregated serving places the \prefill and \decode on isolated workers. However, no single architecture excels in both TTFT and TPOT metrics. After conducting a root cause analysis, we concluded that in \disaggregated LLM serving, \prefill execution has minimal interference with \decode execution but result in high queuing times. In contrast, \nondisaggregated LLM serving effectively reduces queuing times but introduces significant interference between prefills and decodes.

In order to leverage the best aspects of both \nondisaggregated and \disaggregated LLM serving, we have designed and implemented \work. \work introduces an sevice-level objectives (SLO)-aware multiplexing strategy that balances the queuing time and the interference, enabling the LLM serving to achieve high TTFT and TPOT SLOs 
simultaneously. Our evaluation of real-world datasets reveals that \work outperforms both state-of-the-art \nondisaggregated and \disaggregated LLM serving systems, achieving up to $2.09\times$ more requests within a $90\%$ SLO attainment. Specially, compared to the \disaggregated LLM serving system, \work improves P90 TTFT performance by $9\times$ with only an $15\%$ reduction in P90 TPOT. Against the \nondisaggregated LLM serving systems, \work delivers a $2.8\times$ performance improvement in P90 TPOT while maintaining the same P90 TTFT.

\end{abstract}


\section{Introduction}
\label{sec:intro}

With the rapid development of the field of natural language processing, large language models (LLMs)\cite{language-are-few-shot-learners-2020,palm-2024,scaling-laws-2020,Gpt-4-2024, transformer-neurips-2017} have become increasingly versatile.
The transformer based LLM inference can be divided into two phases: the \textit{prefill} and the \textit{decode}.
A request first complete \prefill to generate the first token, and then process the \decode repeatedly, generating a series of tokens in a streaming manner.
In LLM serving, both phases have their own latency metrics that are highly relevant to user experience, which are time-to-first-token (TTFT) for \prefill and time-per-output-token (TPOT) for \decode. TTFT measures the time from when the user sends a request to when they first receive feedback. TPOT describes the average time for a single request to output tokens. As users typically expect to receive feedback quickly, optimizing TTFT is crucial. Besides, if TPOT is too long, users will experience delays while reading or processing the generations of LLM, which will affect the smoothness and continuity of the user experience. For a single request, both the TTFT and TPOT service-level objectives~(SLOs) must be satisfied in LLM Serving.




LLM serving systems are categorized into \nondisaggregated LLM serving systems and \disaggregated LLM serving systems based on whether they separate two phases to isolated workers. \Disaggregated LLM serving~\cite{distserve-osdi-2024, splitwise-isca-2024, mooncake-arxiv-2024, tetriinfer-arxiv-2024} use isolated workers to serve \prefill and \decode dividely. \Nondisaggregated LLM serving systems~\cite{orca-osdi-2022,vllm-sosp-2023, sarathi-serve-osdi-2024,sglang-neurips-2024} maximize GPU resource utilization by colocating \prefill and \decode process in the same worker.
However, no single architecture excels in both TTFT and TPOT SLOs attainment. 
Because of high interference in \nondisaggregated LLM serving and high queuing time in \disaggregated LLM serving. 

\begin{figure}
    \centering
    \includegraphics[width=1\linewidth]{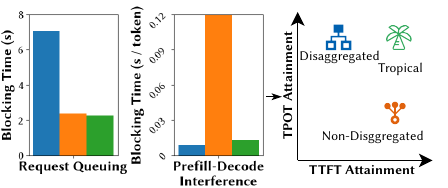}
    \caption{Comparision among \textit{Non-Disaggregated}, \textit{Disaggregated} LLM Serving and \work.}
    \vspace{-2em}
    \label{fig:introduction}
\end{figure}

\Nondisaggregated LLM serving experiences significant interference from \prefill on \decode. We evaluate the interference in \figref{fig:introduction} (b) on a real dataset~\cite{mooncake-arxiv-2024}. We quantize the interference by blocking time, the time a request in \decode is blocked by other \prefill executions. Interference in \nondisaggregated serving is significantly higher than in \disaggregated serving, leading to higer TPOT SLO violations. Instead, \disaggregated LLM serving faces the mismatch between the allocation of \prefill-\decode workers and workload requirements. Because \prefill can only use the resources of the \prefill workers and cannot utilize the idle resources of the \decode workers, the \prefill workers are frequently under high workload, leading to longer queuing times for \prefill.
As shown in \figref{fig:introduction} (a), the P90 queuing time for \disaggregated serving is significantly higher than that for \nondisaggregated serving. 
In \disaggregated LLM serving, dynamically switching the role of \prefill and \decode is one of the strategies to address the issue of excessive \prefill queuing times. However, the overhead of switching from \decode workers to \prefill workers is not negligible. Besides, the short-term volatility and unpredictability of LLM serving workloads also make it difficult for \prefill-\decode worker role switching to keep up with workload changes.

Based on our observations, we have found that collocating \prefill and \decode on the same worker in an SLO-aware manner can also improve the system's SLO attainment without switching overhead, which is called SLO-aware multiplexing. Based on the insight, we have designed and implemented \work. \work is a SLO-aware scheduler that manages multiple workers. \work introduces an multiplexing strategy that effectively reduces queuing time in the scenarios with highly dynamic \prefill workloads. \work keeps the interference within an acceptable range that satisfies both TTFT and TPOT SLOs. As shown in \figref{fig:introduction}, \work achieves a Pareto optimality for TTFT and TPOT SLOs compared with both kinds of LLM serving.

In summary, this paper makes the following contributions:
\begin{enumerate}
    \item We dive into the bottlenecks of \disaggregated and \nondisaggregated LLM serving and reveals that the main reason for the extended TTFT in \disaggregated LLM serving is the queueing time. Conversely, the primary cause of the prolonged TPOT in \nondisaggregated LLM serving is the interference.
    \item We employ SLO-aware multiplexing to balance queuing time and interference, enabling the LLM serving to have high TTFT and TPOT SLOs attainment simultaneously.
    \item Conduct a comprehensive evaluation of \work against state-of-the-art \nondisaggregated and \disaggregated LLM serving.
\end{enumerate}

Our evaluation reveals that \work outperforms both state-of-the-art \nondisaggregated and \disaggregated LLM serving systems, achieving up to 2.09$\times$ more requests within a 90\% SLO attainment. Specially, compared to the \disaggregated LLM serving system, \work improves P90 TTFT by 9$\times$ with only an 15\% reduction in P90 TPOT. Against the \nondisaggregated LLM serving systems, \work delivers a 2.8$\times$ improvement in P90 TPOT while maintaining the same P90 TTFT.

\section{Background}
\label{sec:back}

\subsection{Prefill-Decode Scheduling}
\label{sec:pd_sch}

\begin{figure}
    \centering
    \begin{minipage}[b]{0.99\linewidth} 
        \centering
        \includegraphics[width=\linewidth]{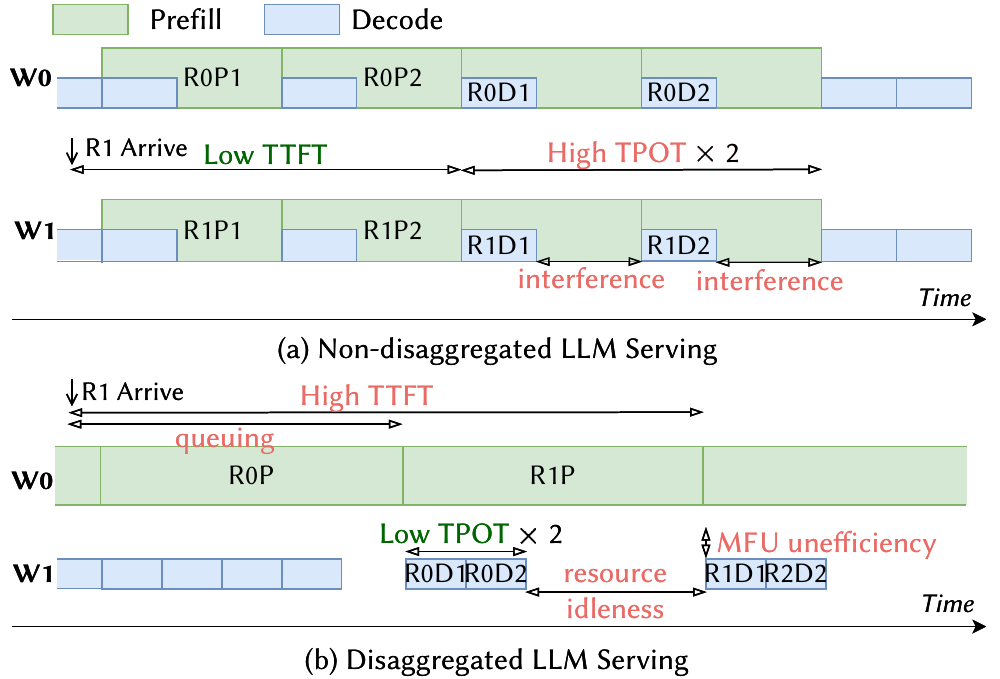}
    \end{minipage}
    \caption{\Prefill-\Decode scheduling comparison.}
    \vspace{-2em}
    \label{fig:iteration}
\end{figure}

Disaggregated LLM serving has longer queuing times, while non-disaggregated LLM serving has more significant interference. We have drawn a timeline example to illustrate the reasons. \figref{fig:iteration} (a) represents an \nondisaggregated serving system~\cite{sarathi-serve-osdi-2024}, where prefills are scheduled eagerly once there are prefills in queue, resulting in a shorter queuing times, which means a shorter TTFT. However, the execution time for prefill is generally longer than that for decode. The execution of prefill will block the execution of decode, leading to a worse TPOT.
State-of-the-art \nondisaggregated LLM serving decompose long prefill into chunks. However, the execution of decode is still slow down because of the interference. Especially, the execution time of the chunks will significantly rise in the long context scenarios.
 \figref{fig:iteration} (b) represents a \disaggregated serving system, where requests in the \prefill phase and the \decode phase are executed on isolated workers. Requests in the 
\decode phase are not interfered with by the other requests in the \prefill phase, thus achieving a better TPOT. However, \prefill can only utilize the resources of one worker, increasing queuing times and resulting in a worse TTFT. In addition, as shown in \figref{fig:iteration} (b), the slower processing speed of the prefill phase also leads to idle time in the decode phase. Besides, since requests in the \prefill and \decode phases cannot be executed in the same batch, the computational resources of the GPU are not fully utilized, leading to \emph{model FLOPs Utilization (MFU) inefficiency}.

\subsection{SLO in LLM Serving}\label{sec:back-alias}

In this section, we define the TTFT SLO, TPOT SLO, and SLO attainment in the LLM serving. In addition, we define \textit{slack} to describe the difference between the actual execution time and the SLO.

Generally, we define TTFT as the time from when the user sends a request to when they first receive feedback. 
Specially, we define TPOT as the total time spent generating tokens divided by the number of tokens generated, which equates to the average delay between tokens. We posit that users will only perceive waiting when the average generation rate surpasses their reading speed. Consider the following example: a LLM serving system rapidly delivers 20 tokens to the user, then encounters a 1-second pause before resuming token delivery. Although the maximum token-to-token delay for this request is 1 second, users still have a sufficient number of tokens to read during the pause, thus maintaining a positive user experience. Conversely, if users do not have enough information to process, they will perceive a lag in response. In \nondisaggregated LLM serving, the TPOT can be break down to the execution time and the interfernce caused by prefill execution. In \disaggregated LLM serving, as the \prefill and \decode are executed in different workers, the interference of \disaggregated can be composed into the decode execution and the migration time of KVCache from other prefill workers.

We define \textit{slack} as the gap between system performance and its SLO. When the system's performance metrics exceed the SLO, this additional performance can be considered as slack, which provides a certain buffer, allowing the system to withstand a certain degree of load changes or performance fluctuations without violating the SLO. We utilize the slack during the \decode phase to handle tasks from \prefill phase without affecting the user experience.

Let $\mathcal{R}$ be the set of total requests $r$, $L_r$ be the total generation lengths of $r$. The request SLO Attainment $A$ is given by the following formulas:
 
\begin{equation}
\mathcal{R}_{TTFT}=\left\{T_{prefill}^{r} \leq TTFT\_SLO\ |\ r \in \mathcal{R}\right\},
\end{equation}

\begin{equation}
\mathcal{R}_{TPOT}=\left\{\frac{\sum{T_{decode}^r}}{L_r} \leq TPOT\_{SLO}\ |\ r \in \mathcal{R}\right\},
\end{equation}

\begin{equation}
A = \frac{|\mathcal{R}_{TTFT} \cap \mathcal{R}_{TPOT}|}{\mathcal{R}}.
\end{equation}

\section{Characterization and Motivation}
\label{sec:motivation}

In this section, we investigate the characteristics of LLM serving to gain key insights and motivations that will guide our design and implementation of \work.

\vspace{-0.5em}
\subsection{Workloads in LLM Serving}
\label{sec:stochastic_process}

\begin{figure}
    \centering
    \includegraphics[width=1\linewidth]{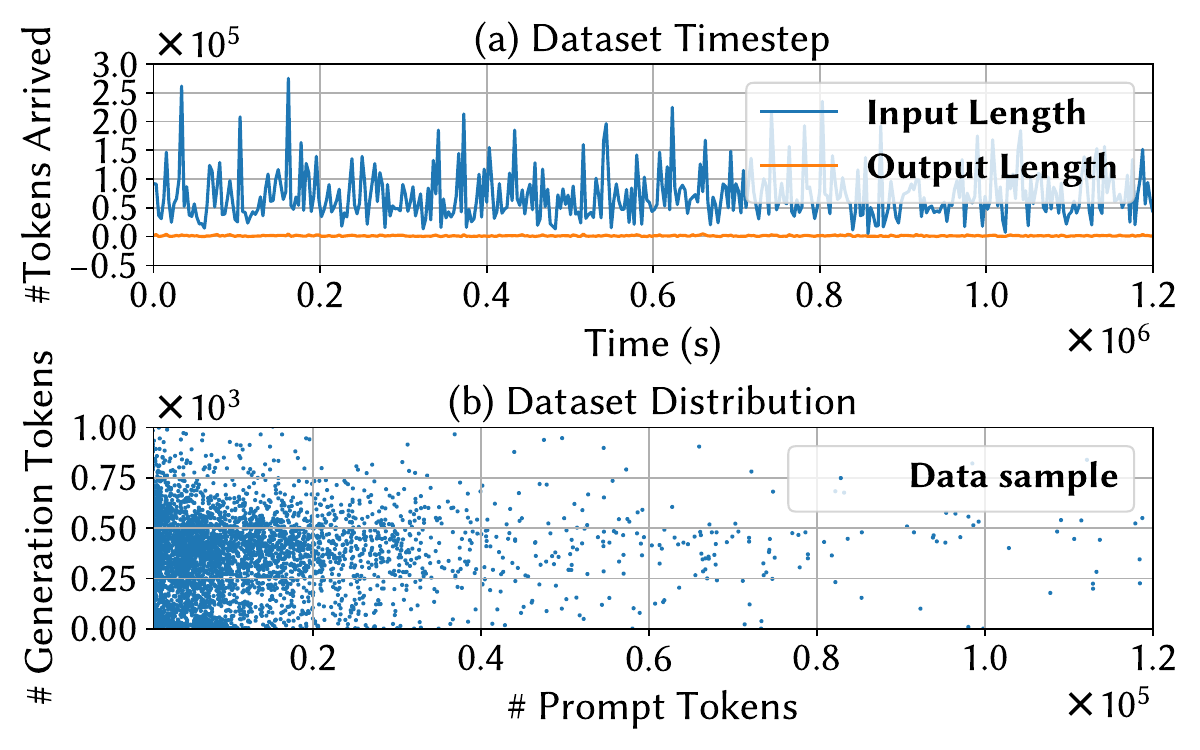}
    \caption{Mooncake Dataset Token Lengths Over Time and Distribution.}
    \label{fig:stochastic_process}
\end{figure}

We observe the dataset~\cite{mooncake-arxiv-2024} to analyze the workload of real-world LLM serving. ~\figref{fig:stochastic_process} (a) shows the total number of tokens arrived during an observation gap. It is evident that the arrival of tokens exhibits extremely high uncertainty in the short term. This uncertainty arises from two main factors. 
First, similar to other DNN serving~\cite{shepherd-osdi-2023, serverlessinthewild-atc-2020}, the serving process itself is random, and user arrivals are unpredictable. 
Additionally, in the case of LLM serving, the number of input tokens for LLM serving also demonstrates significant dynamics. In ~\figref{fig:stochastic_process} (b), we present a scatter plot of a single request's (prefill, decode) sample, which illustrates the input uncertainty. In addition, it can be observed that the distribution of prefill text lengths follows a long-tail pattern. Both \figref{fig:stochastic_process} (a) and (b) indicate that, compared to the output, the input has a higher dynamic range during the LLM serving process. Similar conclusions can also be drawn from the dataset evaluated in other LLM serving-related works~\cite{splitwise-isca-2024}.

\begin{figure}
    \centering
     \includegraphics[width=1\linewidth]{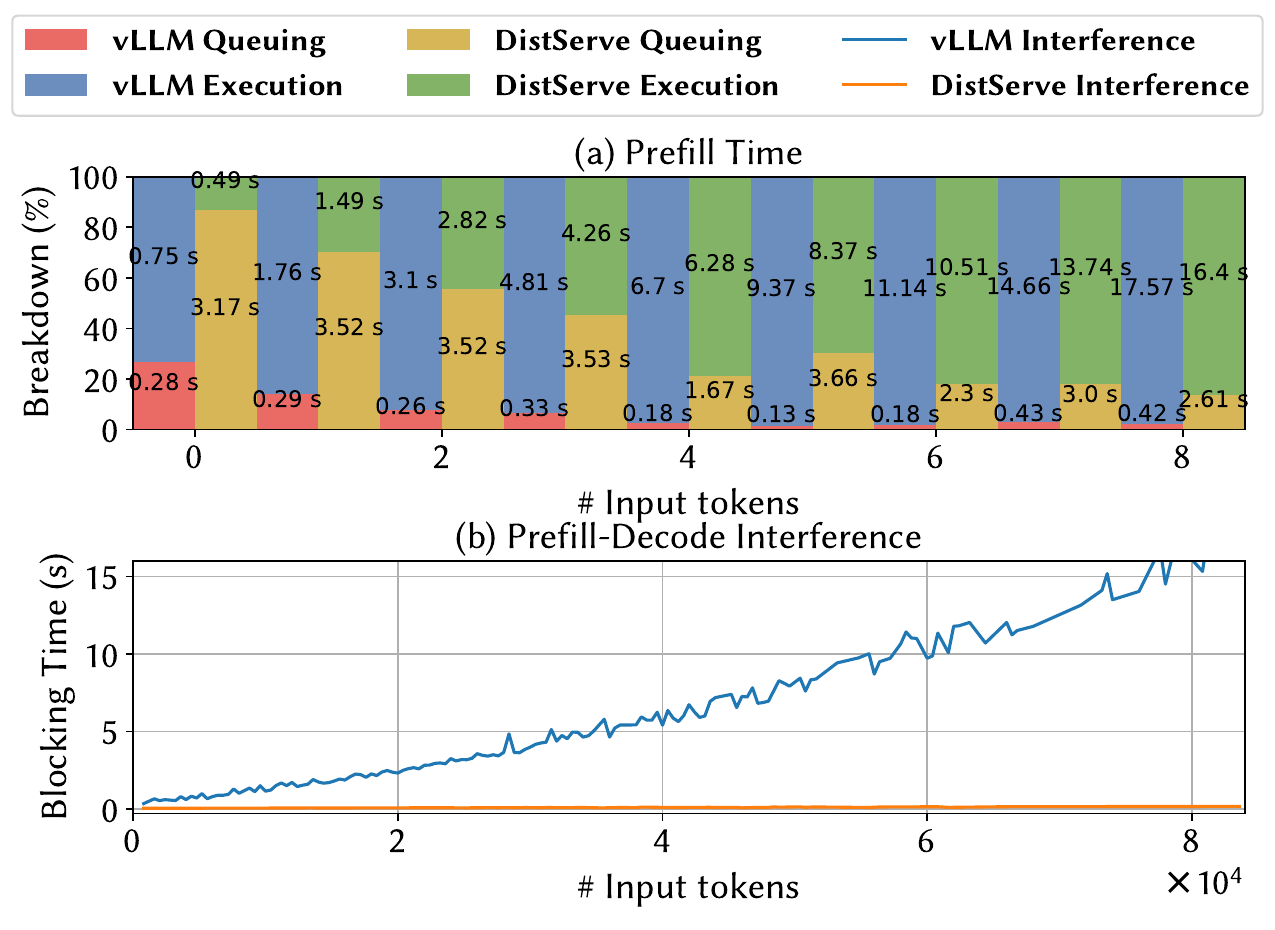}
    \caption{Queuing and Interference Comparision of Non-Disaggregated Serving and Disaggregated Serving.}
    \vspace{-2em}
    \label{fig:non_disagg_vs_disagg}
\end{figure}

\newcounter{mycounter1}
\setcounter{mycounter1}{1} 
\noindent \textbf{Characterization \Roman{mycounter1}:} The lengths of prefill workloads in LLM serving exhibit high dynamism.

\subsection {Efficiency of \textit{Non-Disaggregated} and \textit{Disaggregated} Serving}
\label{sec:nondisagg_or_disagg}
we conduct a quantitative analysis of both \nondisaggregated and \disaggregated scheduling strategies. In \figref{fig:non_disagg_vs_disagg} (a), we analyze the prefill time of the dataset~\cite{mooncake-arxiv-2024} and provided a breakdown to observe the execution time, queuing time, and their proportions for different context lengths. We observed that when the text is short, the TTFT of the \nondisaggregated LLM serving system is significantly better than that of the disaggregated LLM serving system. 
Although the decode phase in \nondisaggregated LLM serving does interfere to the prefill, resulting in slightly higher execution times compared to \disaggregated LLM serving, the queuing time in \nondisaggregated LLM serving is much lower than that in \disaggregated serving. This is particularly noticeable in scenarios with relatively shorter contexts, where the prefill time is dominated by queuing time, which is nearly $10\times$ longer than the execution time. 

We observed that when the context is long, the TPOT of the non-disaggregated LLM serving system is significantly worse than that of the \disaggregated LLM serving system. ~\figref{fig:non_disagg_vs_disagg} (b) illustrates the interference caused by \prefill execution on \decode phase. As the interference primarily arises from the migration of KVCache in \disaggregated LLM serving and from the prefill execution in \nondisaggregated LLM serving. With the support of high-bandwidth KVCache transmission, the migration latency is significantly lower than that of prefill execution time. Therefore, the interference in \disaggregated LLM serving is much less than that in \nondisaggregated LLM serving. It is also evident that as the text length increases, the effectiveness of \disaggregated LLM Serving in mitigating \prefill-\decode interference becomes more pronounced.

By combining the insights from ~\figref{fig:non_disagg_vs_disagg} (a) and (b), we observe that when the context length is relatively short, the queuing time in \prefill phase significantly larger than the interference to \decode phase. This means that, for short texts, the benefits of \disaggregated LLM serving to mitigate interference are likely outweighed by queuing time.

\newcounter{mycounter2}
\setcounter{mycounter2}{2} 
\noindent \textbf{Characterization \Roman{mycounter2}:} The \disaggregated architecture suffers from high queuing times but benefits from interference elimination. In contrast, the \nondisaggregated architecture has lower queuing times but suffers from interference.  Notably, as the length increases, the interference effects in the \nondisaggregated architecture become more pronounced, while for shorter lengths, the proportion of queuing in the \disaggregated architecture dominates.

\begin{figure}
    \centering
     \includegraphics[width=1\linewidth]{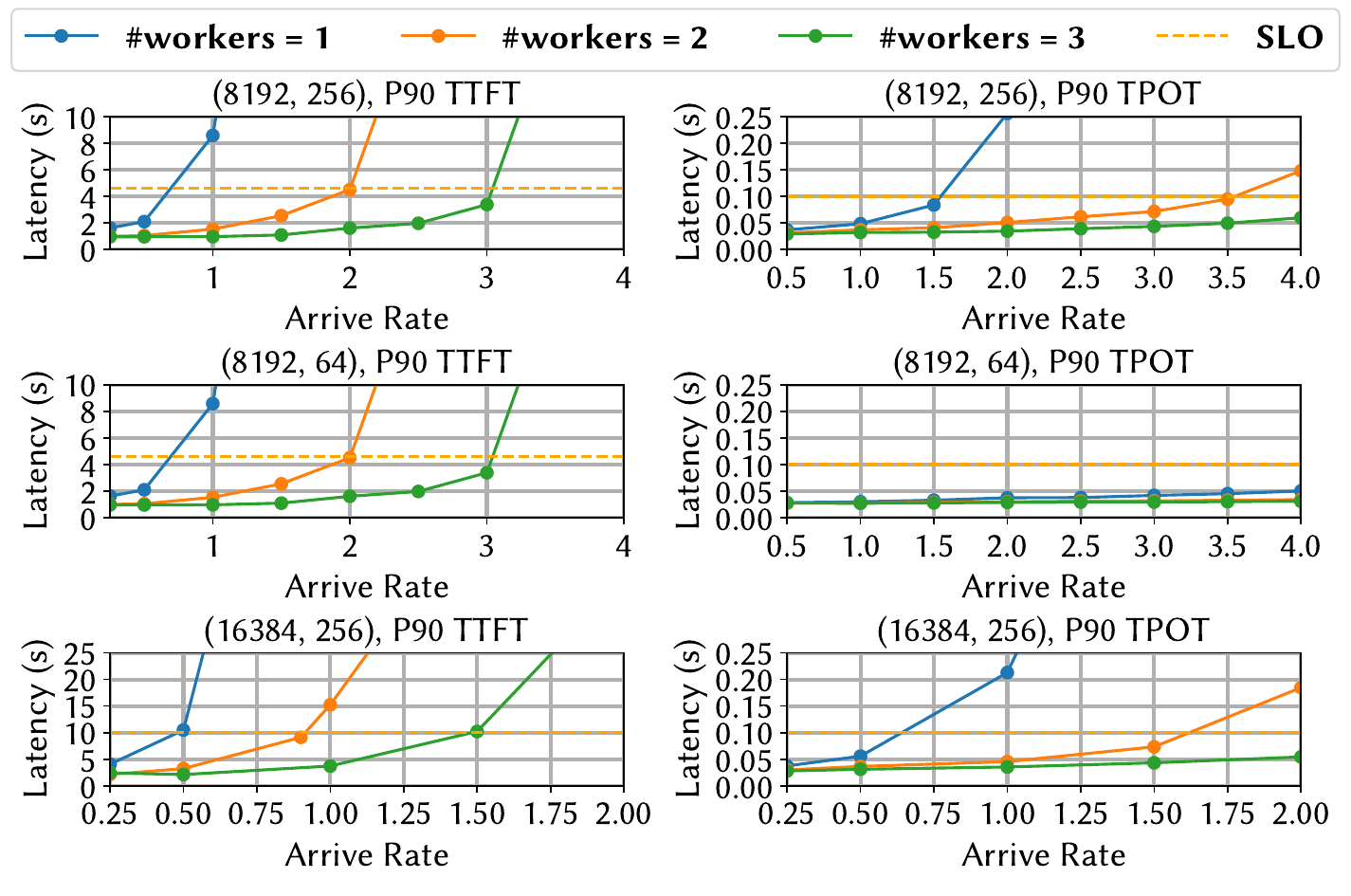}
    \caption{P90 Latency in Different Worker Resource Allocations.}
    \vspace{-2em}
    \label{fig:instance_ability}
\end{figure}

\subsection {Worker Resource Allocation in Disaggregated Serving.}
\label{serving_intensity}
With 4 worker resources, we examine different ratios of worker allocation under three (input, output) length configurations. \figref{fig:instance_ability} shows the performance of P90 latency. We observed that it is difficult to match workloads and resources between \prefill workers and \decode workers simultaneously. With a configuration of (8192, 256), if we allocate 3 workers for \prefill and 1 worker for \decode, the LLM Serving can handle arrival rates of 3.0 for \prefill and 1.6 for \decode. In contrast, with 2 workers for \prefill and 2 for \decode, \prefill can accommodate arrival rates of 2.0, while \decode can handle up to 3.5. 
\figref{fig:instance_ability} also shows the service capability of the LLM is constrained by both the \prefill and \decode capacities. According to the principle of the \emph{leaky bucket effect}, static resource allocation for \prefill and \decode makes it difficult to achieve an optimal match between resources and workloads.
When one phase is overloaded while the other remains idle, the system's performance can deteriorate significantly. 
In addition, due to the higher computational intensity, \prefill requires more computing resources compared to \decode as shown in \figref{fig:instance_ability}. Related works~\cite{splitwise-isca-2024} shows the optimal ratio of \prefill-\decode can be up to 9:1. 
Besides, ~\figref{fig:instance_ability} also reveals that different samples, such as (8192, 64) and (16384, 256), have significantly different optimal ratios of prefill and decode workers.


\newcounter{mycounter3}
\setcounter{mycounter3}{3} 
\noindent \textbf{Characterization \Roman{mycounter3}:} Static ratios of \prefill and \decode worker allocation make it difficult to simultaneously match resources and workloads in both phases.

\vspace{-0.5em}
\subsection {Challenge in LLM Serving}
\label{sec:motivation_challenge}
\vspace{-0.5em}
We summarize the challenges according to the observation.

\noindent\textbf{Queuing and Interference balancing.} From \ref{sec:nondisagg_or_disagg}, it can be observed that \nondisaggregated LLM serving places \prefill and \decode on the same worker. However, the execution of \decode is preempted by \prefill, leading to Interference. 
Besides, if \prefill and \decode are separated and placed on different workers, \prefill can only utilize a portion of the resources, causing \prefill to potentially operate at a high service intensity or even overload, leading to a decline in performance by queuing. Existing LLM serving systems are unable to effectively balance queuing and interference times, making it difficult to simultaneously meet TTFT and TPOT SLOs.

\noindent\textbf{Resource misallocation.} 
According to ~\ref{serving_intensity}, even under fixed input-output length configurations, the ratio of prefill to decode can still lead to resource misallocation. From Section~\ref{sec:stochastic_process}, it can be concluded that the length of requests varies greatly. As a result, the \prefill-\decode ratio can also fluctuate significantly. 
To address the issue of mismatched resource allocation, an effective approach is to implement \prefill-\decode role switching in iteration level~\cite{loongserve-sosp-2024}. The overhead of role switching from \prefill to \decode is minimal, as it only requires requests that have completed the \prefill phase to continue to the \decode phase on the same worker, without the need for inter-worker transmission of KV cache. However, when the system performs the switching from \decode to \prefill, the system will find itself in a dilemma.
Since the \decode phase is stateful, immediately switching a \decode worker to a \prefill worker requires either migration or recomputation, both of which can lead TPOT SLO violations.
If ignored, the \prefill workers will experience overload, leading to the high queuing time. Consequently, achieving a rapid switching from \decode workers to \prefill workers is also quite challenging.

\vspace{-0.5em}
\subsection {Optimization Chance}
\label{sec:observation}
\vspace{-0.3em}
\noindent \textbf{SLO-Aware Multiplexing}. As described in \ref{sec:pd_sch}, We define \textit{slack} as the gap between system performance and its SLO. When the system's performance metrics exceed the SLO, this additional performance can be considered as slack, which provides a certain buffer, allowing the system to withstand a certain degree of workload changes or performance fluctuations without violating the SLO. In LLM serving scenarios, we can utilize the slack during the \decode phase to handle tasks from \prefill phase without affecting the user experience. Whether performing \prefill-\decode \emph{disaggregation} to eliminate interference or colocating \prefill-\decode into the same worker, the goal is to guarantee requests completion within the \emph{SLO}. In a \nondisaggregated LLM serving, \prefill and \decode sharing the same worker is akin to trading interference for reduced queuing time, while disaggregated LLM serving does the opposite. We term the practice of co-locating \prefill and \decode on the same worker while maintaining TTFT and TPOT SLOs as \emph{SLO-aware multiplexing}. If \emph{SLO-aware multiplexing} can be implemented in disaggregated LLM serving, it would be able to reach an equilibrium of queuing time and interference, which enhances the total SLO attainment.

\section{System Design}
\label{sec:sys}

Based on the observations and the solution presented in \secref{sec:motivation}, we design \work, an SLO-aware scheduler to balance the interference and queuing time.

\subsection{Overview}
\label{sec:syc-overview}
\begin{figure}
    \centering
    \includegraphics[width=\linewidth]{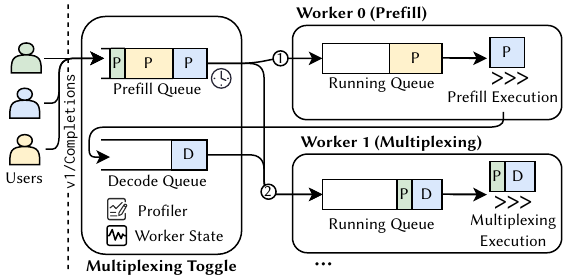}
    \caption{Overview of \work.}
    \vspace{-1em}
    \label{fig:overview}
\end{figure}
The design of \work is illustrated in \figref{fig:overview}. 
\work consists of two parts: the \emph{multiplexing toggle} and a series of \emph{workers}. In \work, workers can perform both \prefill and \decode execution. The phase in which they execute depends on the tasks sent to them by the multiplexing toggle.
The multiplexing toggle primarily performs two functions: assignment and dispatching. For assignment, the multiplexing toggle assigns workers as prefill workers or multiplexing workers.
For dispatching, It places incoming requests into the prefill queue initially and then dispatches them in an SLO-aware manner to competent workers.
We will provide a detailed introduction to the design of the multiplexing toggle
and its internal SLO-aware multiplexing mechanism in \secref{sec:multiplexing_toggle} 
 and \secref{sec:slo_aware_multiplexing}.

\subsection{SLO-Aware Multiplexing}
\label{sec:slo_aware_multiplexing}


As described in Figure \ref{fig:non_disagg_vs_disagg}, the \disaggregated architecture suffers from high queuing times but benefits from interference elimination. In contrast, the \nondisaggregated architecture has lower queuing times but suffers from interference. 
Additionally, as described in \ref{sec:back-alias} and \ref{serving_intensity}, decode workers experience slack due to load fluctuations, and we aim to utilize the slack to serve prefills, thereby reducing the queuing time. 
In order to determine whether the interference from \prefill execution will disrupt the requests that are currently in the \decode phase, we will record the difference between the time taken for a single decode operation and the TPOT time, which is known as the TPOT slack in a single iteration.

\begin{figure}
    \centering
    \includegraphics[width=.93\linewidth]{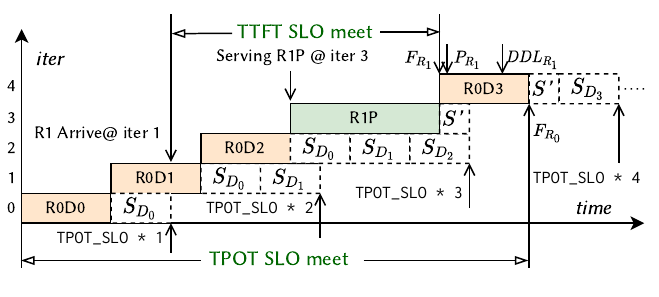}
    \caption{P-D SLO-aware multiplexing by slack checking.}
    \label{fig:pd_multiplexing}
    \vspace{-1em}
\end{figure}

If the predicted execution time is less than the slack time available in the \decode worker, we treat the multiplexing worker as idle and schedule the prefill to the multiplexing worker to alleviate the pressure on the \prefill workers. Although the insertion of \prefill requests may lead to uneven output latency in decoding, recent advancements in LLM serving technologies~\cite{andes-arxiv-2024, llumnix-osdi-2024} aim to reduce fluctuations during the output process, ensuring a smoother flow.
As described in \figref{fig:pd_multiplexing}, the \decode process to generate the third token of $R_0$ has accumulated enough slack to accommodate a prefill request. Therefore, \work routes a request from the prefill queue for execution. Although the prefill is inserted, $R_0$ does not violate the TPOT SLO.

Due to the significantly shorter execution time of short \prefill than the long, it is more likely to route the short \prefill request to multiplexing workers.
As we concluded in \ref{fig:non_disagg_vs_disagg}, short \prefill are queueing dominated, which gives \nondisaggregated LLM serving an advantage. In contrast, long \prefill cause greater interference with decoding, which makes \disaggregated LLM serving outperform. SLO-aware multiplexing leverages the benefits of both serving strategy.

\subsection{Multiplexing Toggle}
\label{sec:multiplexing_toggle}
As described in \figref{fig:overview}, there are two paths for sending requests to workers. The multiplexing toggle provides a traffic control mechanism: when the system is dominated by queuing, it follows Path \normalsize{\textcircled{\scriptsize{1}}}\normalsize, and when interference dominates, it follows Path \normalsize{\textcircled{\scriptsize{2}}}\normalsize.
The multiplexing toggle can dispatch the request through Path \normalsize{\textcircled{\scriptsize{1}}}\normalsize\xspace to \prefill workers for \prefill execution, and then enter to the \decode queue of the multiplexing toggle.
The multiplexing toggle will dispatch the request again through Path \normalsize{\textcircled{\scriptsize{2}}}\normalsize\xspace to the multiplexing worker for \decode execution. 
Alternatively, requests can bypass Path \normalsize{\textcircled{\scriptsize{1}}}\normalsize\xspace and directly enter the multiplexing workers. When the interference from \prefill on \decode remains within acceptable limits as described in \ref{sec:slo_aware_multiplexing}, requests can be sent directly via Path \normalsize{\textcircled{\scriptsize{2}}}\normalsize\xspace.
When the interference is not negligible, the \prefill worker acts as an \emph{interference inhibitor}, reducing the interference to an acceptable range.
Since \prefill can utilize the worker resources of both \prefill and multiplexing workers, the system's queuing time will be significantly less than that of \disaggregated serving. 

The multiplexing toggle records the status of each worker, including monitoring the HBM watermark and the local queue of workers. 
Our design consolidates choice in Path \normalsize{\textcircled{\scriptsize{1}}}\normalsize\xspace and \normalsize{\textcircled{\scriptsize{2}}}\normalsize\xspace. Additionally, since the execution time of the LLM's \prefill phase is highly predictable~\cite{clockwork-osdi-2022}, we leverage offline profiling tools to estimate both the execution time of a \prefill request and the queuing time when scheduling to the local worker.
The multiplexing toggle conservatively sends requests to prefill workers only when the addition of the predicted execution time and the predicted queuing time in the workers, is less than the available slack to TTFT SLO.
When a competent worker is found, an scheduling operator is created and requests are dequeued from \prefill queue and enqueue to the running queue of the local worker.

\work also conservatively sends prefill requests to multiplexing workers. In addition to the TPOT slack budget outlined in ~\ref{sec:slo_aware_multiplexing}, the multiplexing toggle records and predicts the execution time of the decoding batch for multiplexing workers. Given the high predictability of execution times, the multiplexing toggle will not route prefill requests to multiplexing workers when the execution time of the decoding batch is approaching the preset TPOT SLO. Additionally, since HBM usage is a critical factor affecting decoding execution, the multiplexing toggle will not select a multiplexing worker if the HBM usage for decoding exceeds the threshold. Furthermore, for multiplexing workers, we will also utilize chunked prefill to further smooth out the interference that prefill may cause to decoding.

When the watermark of all decode workers exceeds its threshold, indicating insufficient resources for \decode workers, \prefill workers can be directly assigned as multiplexing workers. If prefill frequently encounters delays (such as when \prefill texts are too long), this will also trigger a transition from multiplexing workers back to prefill workers for a higher \prefill throughput.

\section{Evaluation}
\label{sec:eval}

In this section, we evaluate the efficiency of \work.

\subsection{Experimental Setup}
\label{sec:eval-setup}

\noindent \textbf{Models}. We choose InternLM-20B~\cite{internlm2-arxiv-2024}, an LLM that is popular and performs exceptionally well in long context scenarios, the max supported context window is 200K, which satisfies the requirement of the maximum number of tokens in the dataset.

\noindent \textbf{Testbed}. We evaluated \work on a server equipped with 8 NVIDIA A100 80GB GPUs, with 600 GB/s P2P bandwidth between GPUs. 

\noindent \textbf{workloads}. We choose Mooncake~\cite{mooncake-arxiv-2024}. It is obtained from the trace of real long context LLM serving.

\noindent \textbf{Metrics}. We use SLO attainment as the major evaluation metric. Same as previous relative LLM serving works~\cite{splitwise-isca-2024, fastserve-arxiv-2024}. We set the TTFT SLO to $5 \times$ the execution time of the latency in the corresponding phase under light worload. The other metrics we are concerned with are TTFT and TPOT average and P90 latency. To find the sources of benefit for TTFT, we analyze the queuing time. And we also show the cumulative distribution function (CDF) of the TTFT and TPOT.

\noindent \textbf{Baselines}. We compare \work with the following LLM serving:
\begin{itemize}
    \item \textbf{vLLM}~\cite{vllm-sosp-2023} vLLM is a representative \nondisaggregated LLM serving system widely used in both academia and industry. It colocates the \prefill and \decode computation at the same worker and struggles to meet TPOT SLO.
    \item \textbf{vLLM with chunked prefill~\cite{sarathi-serve-osdi-2024}}. Breaking down the prefill phase into multiple smaller chunks, and executing only one prefill chunk per iteration~\cite{sarathi-serve-osdi-2024}. This approach can shorten the time prefills interfere decodes. However, this method still cannot effectively control the interference of prefills on decodes. We set the chunk size to 2048 based on the profiling of the workload.
    \item \textbf{DistServe}~\cite{distserve-osdi-2024}. We implemented \disaggregated serving based on vLLM~\cite{vllm-sosp-2023}, executing prefills and decodes in isolated workers, which is consistent with the implementation in DistServe~\cite{distserve-osdi-2024}.
\end{itemize}

Both the baselines and \work divided the 8 GPUs into 4 workers, configuring the degree of tensor parallelism of each worker to be 2. To evaluate the efficiency of SLO-aware multiplexing, \work and DistServe all use 2 wrokers for \prefill and 2 workers for \decode.

For both baselines and \work, we adapt the dispatching strategy of InFaas~\cite{infaas-atc-2021} with LLM serving, where the global dispatcher sends requests to the workers with the fewest unfinished token counts. In \work, the multiplexing toggle keeps track of the slack budget for multiplexing workers. Under the condition of satisfying the SLO, the multiplexing toggle will trigger multiplexing to piggyback \prefill with the \decode requests for execution.

\vspace{-0.5em}
\subsection{SLO Attainment}
\begin{figure}
    \centering
    \begin{minipage}[b]{\linewidth} 
        \centering
        \includegraphics[width=\linewidth]{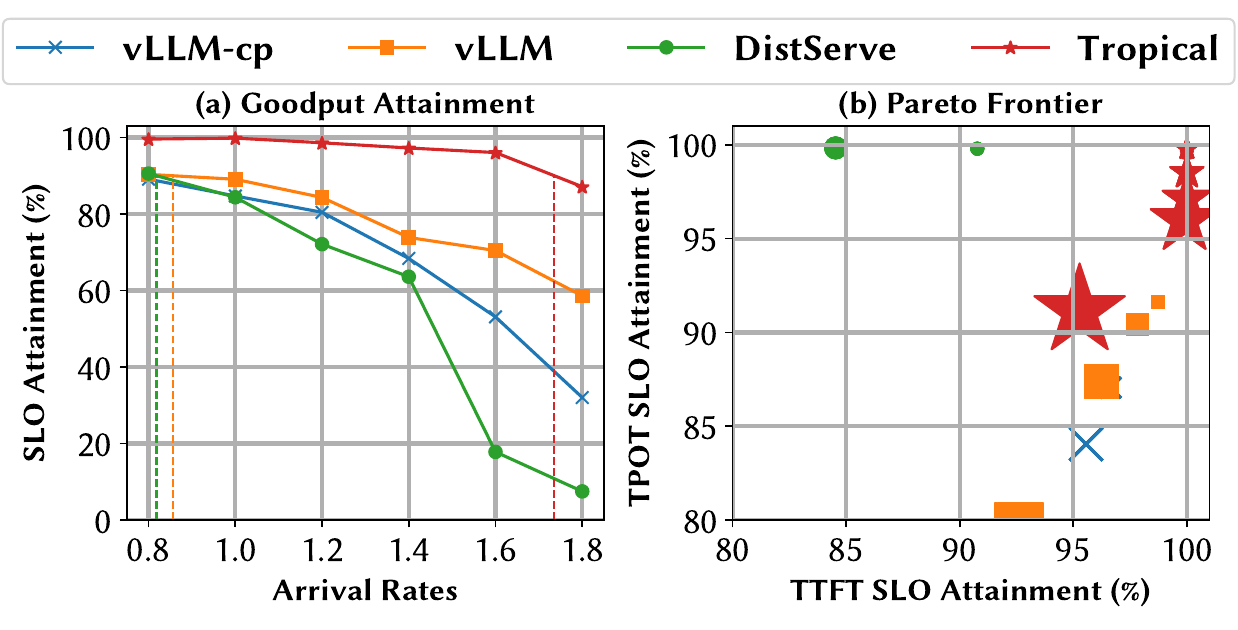}
    \end{minipage}
    \caption{SLO attainment.}
    \label{fig:mooncake_slo_20B}
    \vspace{-1em}
\end{figure}

\vspace{-0.3em}
In this section, we compare the end-to-end preformance of \work against the baselines on real world datasets. \figref{fig:mooncake_slo_20B} (a) illustrates that when we gradually increase the rate, more requests will violate the latency requirements and the SLO attainment decreases. Compared to baselines, \work has improved performance by serving 2.02x more users. From  \figref{fig:mooncake_slo_20B}, it can be observed that the performance of chunked-prefill is not optimal. This is due to the fact that as the chunk index increases, the time to compute a chunk also increases, which can probably lead to TPOT SLO violation. \nondisaggregated LLM serving also has lower SLO attainment than \work. To evaluate that the SLO violation is caused by queuing delay or interference.
We further break down the SLO violation into TTFT and TPOT violation, we show the Pareto frontier diagram in \figref{fig:mooncake_slo_20B} (b), analyzing the changes in TTFT SLO attainment and TPOT SLO attainment for different LLM serving systems with the arrival rate. We found that DistServe has worse TTFT SLO Attainment than others, indicating that the performance of \disaggregated LLM serving is dominated by the queuing time in the prefill phase, while vLLM has higher TTFT SLO attainment, but compared to DistServe, its TPOT SLO is significantly reduced, suggesting that \nondisaggregated LLM serving is dominated by \prefill-\decode interference. Evaluation show that the SLO-aware multiplexing can significantly reduce the TTFT queuing time and interference to ensure both TTFT and TPOT SLOs attainment.

\begin{figure}
    \centering
    \begin{minipage}[b]{\linewidth} 
        \centering
        \includegraphics[width=\linewidth]{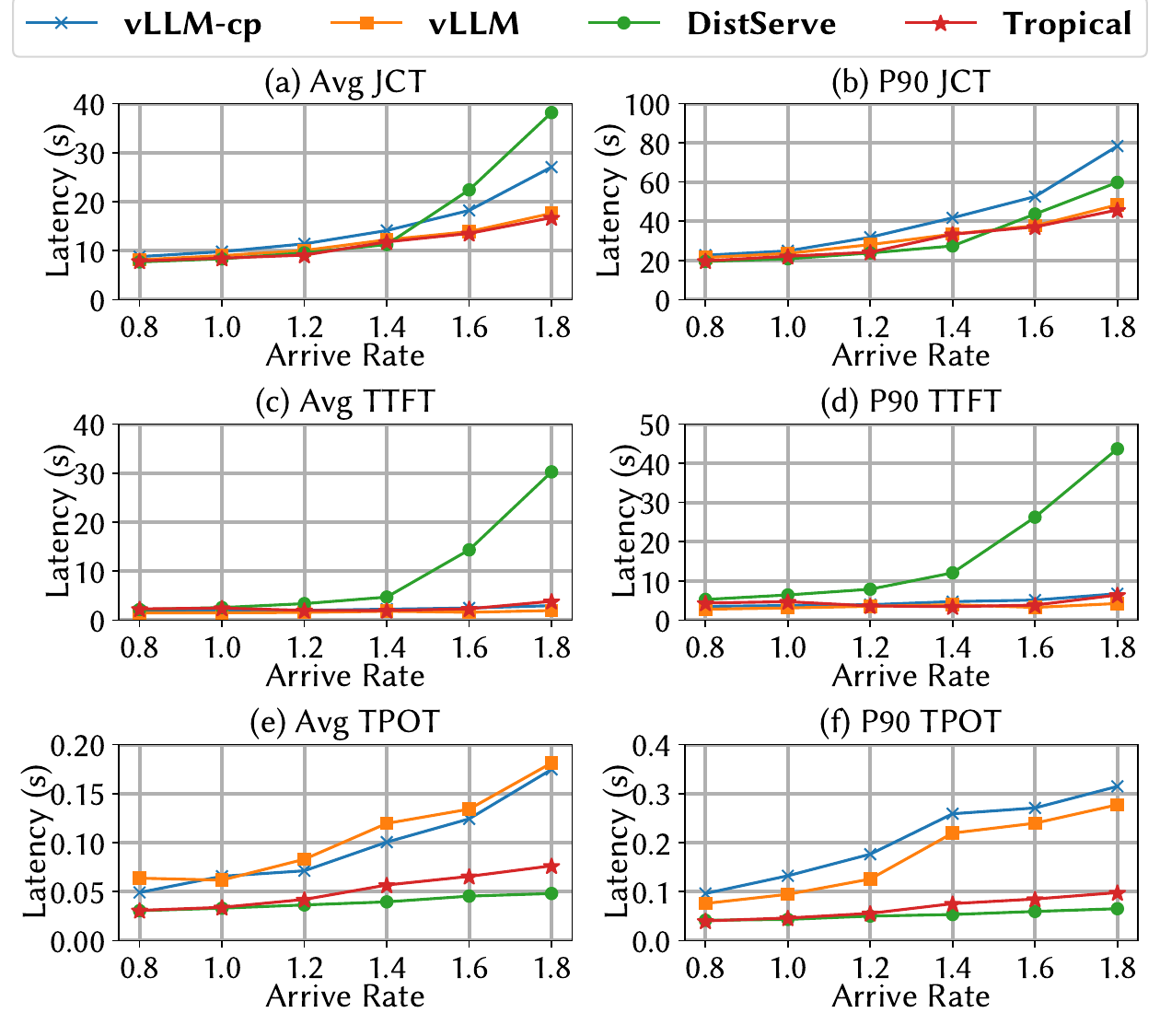}
    \end{minipage}
    \caption{Average and P90 Latency.}
    \label{fig:mooncake_latency_20B}
    \vspace{-2em}
\end{figure}

\vspace{-0.4em}
\subsection{Latency}
\vspace{-0.6em}
As shown in \figref{fig:mooncake_latency_20B} (a) and (b), \work and vLLM outperform the other two serving systems for total latency. vLLM improves system worker resource utilization by co-location. \work achieves high worker resource utilization through SLO-aware multiplexing. \work improves TTFT metric by 9× compared to DistServe.  
In ~\ref{fig:mooncake_latency_20B} (e) and (f), the TPOT performance of \work is up to $2.33\times$ against vLLM and vLLM w/ chunked prefill. The TPOT performance of \work is slightly lower than DistServe. However, \work still meets the TPOT SLO for most requests as shown in \figref{fig:mooncake_slo_20B}.

\begin{figure}
    \centering
    \begin{minipage}[b]{\linewidth} 
        \centering
        \includegraphics[width=\linewidth]{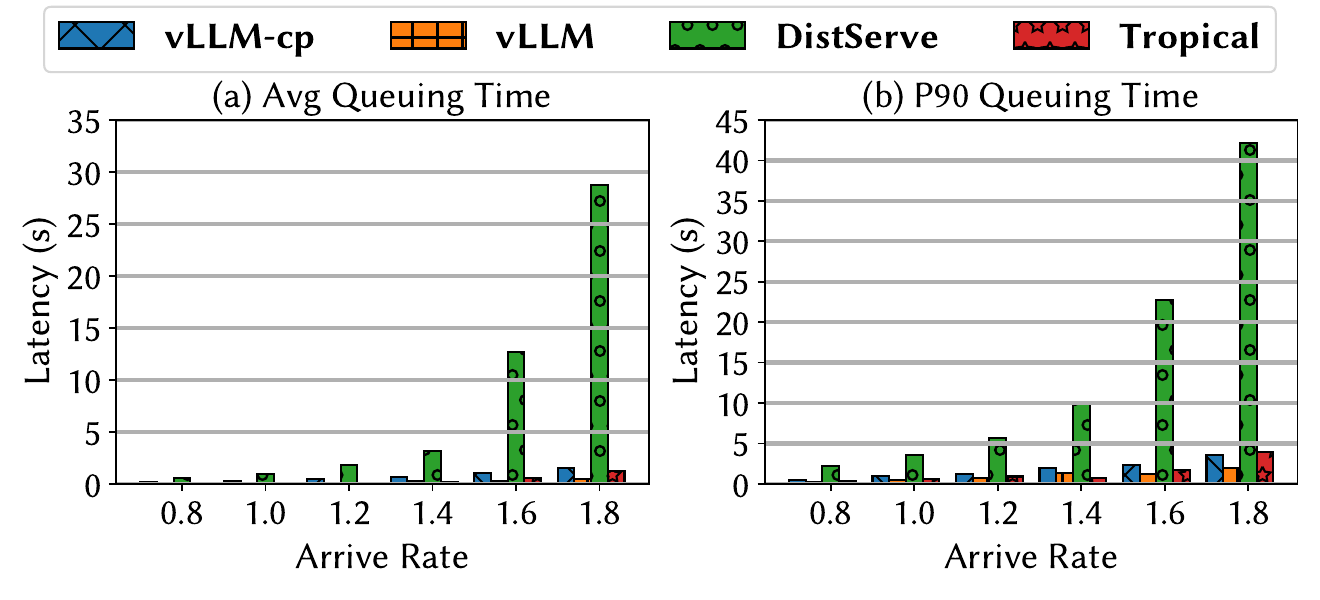}
    \end{minipage}
    \caption{Queuing Time.}
    \vspace{-1em}
    \label{fig:mooncake_queuing_20B}
\end{figure}

\begin{figure}
    \centering
    \begin{minipage}[b]{\linewidth} 
        \centering
        \includegraphics[width=\linewidth]{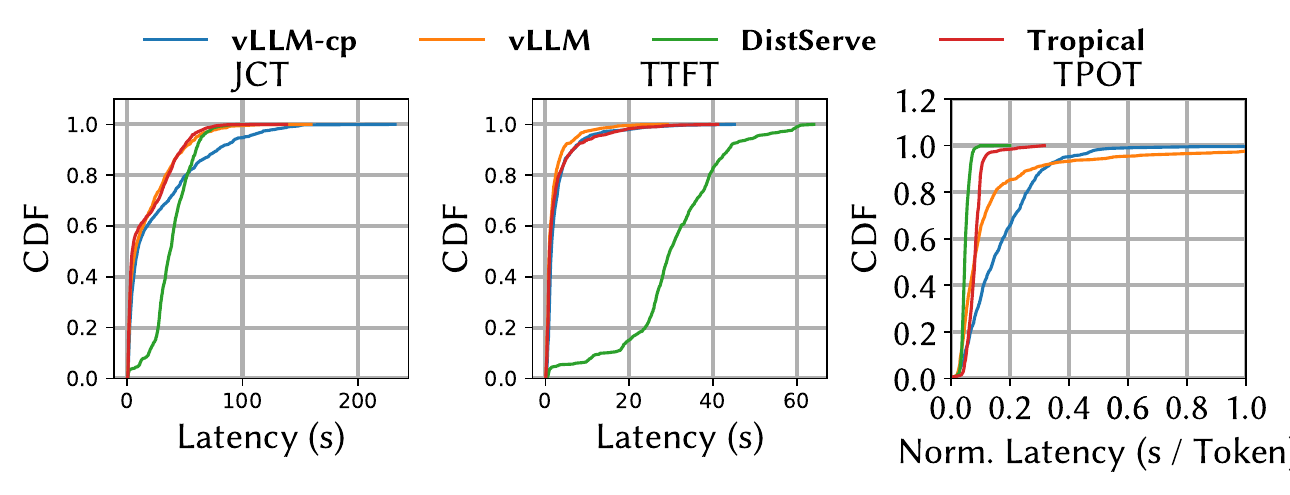}
    \end{minipage}
    \vspace{-1em}
    \caption{CDF.}
    \label{fig:cdf}
    \vspace{-2em}
\end{figure}

\vspace{-1em}
\subsection{Queueing Time}
\vspace{-0.3em}
To further analyze the sources of TTFT performance optimization against \disaggregated LLM serving, we profile the queuing times of different LLM serving systems, as shown in \figref{fig:mooncake_queuing_20B}. Tropical improves P90 queueing time by 9× compared to DistServe. The average queuing time and P90 queuing time of \work are slight higher than vLLM in high arrival rates. However, as shown in ~\figref{fig:non_disagg_vs_disagg}, We use prefill workers to reduce the \decode-\prefill interference, therefore, the TTFT for \work and vLLM are essentially the same.

\vspace{-0.5em}
\subsection{CDF}
\vspace{-0.3em}
The CDF of serving is shown in \figref{fig:cdf}. It can be observed that, except for \work, all other serving systems exhibit significant tail latency. Instead, \work achieves a balance between TTFT and TPOT, thereby improving the overall service quality of the system.

\vspace{-0.5em}
\section{Related Work}
\vspace{-0.5em}
\label{sec:related}

\noindent \textbf{LLM Serving.} The \nondisaggregated inference service systems~\cite{muxserve-icml-2025,vllm-sosp-2023,sglang-neurips-2024,orca-osdi-2022,aplaserve-osdi-2023}, which process the prefill phase and decode phase within the same worker. The \disaggregated serving system~\cite{distserve-osdi-2024,splitwise-isca-2024,mooncake-arxiv-2024, spotserve-asplos-2024,exegpt-asplos-2024, pdserve-arxiv-2024, dejavu-arxiv-2024, memserve-arxiv-2024} handle the prefill phase and decode phase using different workers, eliminating the interference between requests.
In the field of LLM serving, kernel optimization is a critical area of research. These optimization techniques demonstrate the diversity and depth of kernel optimization in LLM serving\cite{flashattention-2022, flashdecoding++-mlsys-2023}. In the context of Deep Neural Networks (DNNs), there are numerous SLO-aware scheduling strategies~\cite{shepherd-osdi-2023,infaas-atc-2021,clockwork-osdi-2022, reef-osdi-2022, grandslam-eurosys-2019}. Andes~\cite{andes-arxiv-2024} defines a Quality of Experience (QoE) metric and enhanced the service quality for users in the LLM serving scenario by optimizing this metric. DistServe~\cite{distserve-osdi-2024} introduced the concept of Goodput and defined the TTFT SLO and TPOT SLO for the LLM serving service scenario. However, it did not consider maintaining a high resource utilization while simultaneously meeting TTFT and TPOT SLOs by SLO-aware multiplexing.


\section{Conclusion}

We conducted a detailed analysis of the drawback of existing \disaggregated LLM serving  in TTFT by queuing time, and the deficiencies of existing \nondisaggregated LLM serving systems in TPOT by interference. 
We design \work, a multi-worker scheduler. \work balances queuing time and interference by SLO-aware multiplexing, enabling the LLM serving to have high TTFT and TPOT SLOs attainment simultaneously.

\section*{Acknowledgement}
\vspace{-0.3em}

The authors sincerely thank anonymous DAC reviewers for their valuable comments on this paper. The research is supported by the Shanghai Municipal Science and Technology Major Project. Greatly thank Zihan Wang for his insightful feedback and constructive suggestions, which significantly improved the quality and conceptual framework of this work.


\bibliography{bibfile}

@inproceedings {orca-osdi-2022,
author = {Gyeong-In Yu and others},
title = {Orca: A Distributed Serving System for {Transformer-Based} Generative Models},
booktitle = {16th USENIX Symposium on Operating Systems Design and Implementation (OSDI 22)},
year = {2022},
isbn = {978-1-939133-28-1},
address = {Carlsbad, CA},
pages = {521--538},
url = {https://www.usenix.org/conference/osdi22/presentation/yu},
publisher = {USENIX Association},
month = jul
}

@article{Gpt-4-2024,
  title={Gpt-4 technical report},
  author={Achiam, Josh and Adler, Steven and Agarwal, Sandhini and Ahmad, Lama and Akkaya, Ilge and Aleman, Florencia Leoni and Almeida, Diogo and Altenschmidt, Janko and Altman, Sam and Anadkat, Shyamal and others},
  journal={arXiv preprint arXiv:2303.08774},
  year={2023}
}

@article{flashattention-2022,
  title={Flashattention: Fast and memory-efficient exact attention with io-awareness},
  author={Dao, Tri and others},
  journal={Advances in Neural Information Processing Systems},
  volume={35},
  pages={16344--16359},
  year={2022}
}

@article{flashdecoding++-mlsys-2023,
  title={Flashdecoding++: Faster large language model inference on gpus},
  author={Hong, Ke and others},
  journal={arXiv preprint arXiv:2311.01282},
  year={2023}
}

@inproceedings{transformer-neurips-2017,
author = {Vaswani and others},
title = {Attention is all you need},
year = {2017},
isbn = {9781510860964},
publisher = {Curran Associates Inc.},
address = {Red Hook, NY, USA},
booktitle = {Proceedings of the 31st International Conference on Neural Information Processing Systems},
location = {Long Beach, California, USA},
series = {NIPS'17}
}

@inproceedings{language-are-few-shot-learners-2020,
author = {Brown and others},
title = {Language models are few-shot learners},
year = {2020},
isbn = {9781713829546},
publisher = {Curran Associates Inc.},
address = {Red Hook, NY, USA},
booktitle = {Proceedings of the 34th International Conference on Neural Information Processing Systems},
articleno = {159},
numpages = {25},
location = {Vancouver, BC, Canada},
series = {NIPS '20}
}

@article{scaling-laws-2020,
  title={Scaling laws for neural language models},
  author={Kaplan, Jared and McCandlish, Sam and Henighan, Tom and Brown, Tom B and Chess, Benjamin and Child, Rewon and Gray, Scott and Radford, Alec and Wu, Jeffrey and Amodei, Dario},
  journal={arXiv preprint arXiv:2001.08361},
  year={2020}
}

@article{palm-2024,
author = {Chowdhery, Aakanksha and others},
title = {PaLM: scaling language modeling with pathways},
year = {2024},
issue_date = {January 2023},
publisher = {JMLR.org},
volume = {24},
number = {1},
issn = {1532-4435},
journal = {J. Mach. Learn. Res.},
month = mar,
articleno = {240},
numpages = {113}
}

@inproceedings{vllm-sosp-2023,
author = {Kwon and others},
title = {Efficient Memory Management for Large Language Model Serving with PagedAttention},
year = {2023},
isbn = {9798400702297},
publisher = {Association for Computing Machinery},
address = {New York, NY, USA},
url = {https://doi.org/10.1145/3600006.3613165},
doi = {10.1145/3600006.3613165},
booktitle = {Proceedings of the 29th Symposium on Operating Systems Principles},
pages = {611–626},
numpages = {16},
location = {Koblenz, Germany},
series = {SOSP '23}
}

@inproceedings {sarathi-serve-osdi-2024,
author = {Amey Agrawal and others},
title = {Taming {Throughput-Latency} Tradeoff in {LLM} Inference with {Sarathi-Serve}},
booktitle = {18th USENIX Symposium on Operating Systems Design and Implementation (OSDI 24)},
year = {2024},
isbn = {978-1-939133-40-3},
address = {Santa Clara, CA},
pages = {117--134},
url = {https://www.usenix.org/conference/osdi24/presentation/agrawal},
publisher = {USENIX Association},
month = jul
}

@inproceedings {distserve-osdi-2024,
author = {Yinmin Zhong and others},
title = {{DistServe}: Disaggregating Prefill and Decoding for Goodput-optimized Large Language Model Serving},
booktitle = {18th USENIX Symposium on Operating Systems Design and Implementation (OSDI 24)},
year = {2024},
isbn = {978-1-939133-40-3},
address = {Santa Clara, CA},
pages = {193--210},
url = {https://www.usenix.org/conference/osdi24/presentation/zhong-yinmin},
publisher = {USENIX Association},
month = jul
}

@INPROCEEDINGS{splitwise-isca-2024,
  author={Patel, Pratyush and others},
  booktitle={2024 ACM/IEEE 51st Annual International Symposium on Computer Architecture (ISCA)}, 
  title={Splitwise: Efficient Generative LLM Inference Using Phase Splitting}, 
  year={2024},
  volume={},
  number={},
  pages={118-132},
  doi={10.1109/ISCA59077.2024.00019}}

@misc{tetriinfer-arxiv-2024,
      title={Inference without Interference: Disaggregate LLM Inference for Mixed Downstream Workloads}, 
      author={Cunchen Hu and others},
      year={2024},
      eprint={2401.11181},
      archivePrefix={arXiv},
      primaryClass={cs.DC},
      url={https://arxiv.org/abs/2401.11181}, 
}

@misc{mooncake-arxiv-2024,
      title={Mooncake: A KVCache-centric Disaggregated Architecture for LLM Serving}, 
      author={Ruoyu Qin and others},
      year={2024},
      eprint={2407.00079},
      archivePrefix={arXiv},
      primaryClass={cs.DC},
      url={https://arxiv.org/abs/2407.00079}, 
}

@inproceedings {reef-osdi-2022,
author = {Mingcong Han and others},
title = {Microsecond-scale Preemption for Concurrent {GPU-accelerated} {DNN} Inferences},
booktitle = {16th USENIX Symposium on Operating Systems Design and Implementation (OSDI 22)},
year = {2022},
isbn = {978-1-939133-28-1},
address = {Carlsbad, CA},
pages = {539--558},
url = {https://www.usenix.org/conference/osdi22/presentation/han},
publisher = {USENIX Association},
month = jul
}

@misc{andes-arxiv-2024,
      title={Andes: Defining and Enhancing Quality-of-Experience in LLM-Based Text Streaming Services}, 
      author={Jiachen Liu and others},
      year={2024},
      eprint={2404.16283},
      archivePrefix={arXiv},
      primaryClass={cs.DC},
      url={https://arxiv.org/abs/2404.16283}, 
}

@misc{fastserve-arxiv-2024,
      title={Fast Distributed Inference Serving for Large Language Models}, 
      author={Bingyang Wu and others},
      year={2024},
      eprint={2305.05920},
      archivePrefix={arXiv},
      primaryClass={cs.LG},
      url={https://arxiv.org/abs/2305.05920}, 
}

@misc{internlm2-arxiv-2024,
      title={InternLM2 Technical Report},
      author={Zheng Cai and others},
      year={2024},
      eprint={2403.17297},
      archivePrefix={arXiv},
      primaryClass={cs.CL}
}

@inproceedings{muxserve-icml-2025,
title={MuxServe: Flexible Spatial-Temporal Multiplexing for Multiple {LLM} Serving},
author={Jiangfei Duan and others},
booktitle={Forty-first International Conference on Machine Learning},
year={2024},
url={https://openreview.net/forum?id=R0SoZvqXyQ}
}

@inproceedings{sglang-neurips-2024,
title={{SGL}ang: Efficient Execution of Structured Language Model Programs},
author={Lianmin Zheng and others},
booktitle={The Thirty-eighth Annual Conference on Neural Information Processing Systems},
year={2024},
url={https://openreview.net/forum?id=VqkAKQibpq}
}

@inproceedings {shepherd-osdi-2023,
author = {Hong Zhang and others},
title = {{SHEPHERD}: Serving {DNNs} in the Wild},
booktitle = {20th USENIX Symposium on Networked Systems Design and Implementation (NSDI 23)},
year = {2023},
isbn = {978-1-939133-33-5},
address = {Boston, MA},
pages = {787--808},
url = {https://www.usenix.org/conference/nsdi23/presentation/zhang-hong},
publisher = {USENIX Association},
month = apr
}

@inproceedings {clockwork-osdi-2022,
author = {Arpan Gujarati and others},
title = {Serving {DNNs} like Clockwork: Performance Predictability from the Bottom Up},
booktitle = {14th USENIX Symposium on Operating Systems Design and Implementation (OSDI 20)},
year = {2020},
isbn = {978-1-939133-19-9},
pages = {443--462},
url = {https://www.usenix.org/conference/osdi20/presentation/gujarati},
publisher = {USENIX Association},
month = nov
}

@inproceedings {infaas-atc-2021,
author = {Francisco Romero and others},
title = {{INFaaS}: Automated Model-less Inference Serving},
booktitle = {2021 USENIX Annual Technical Conference (USENIX ATC 21)},
year = {2021},
isbn = {978-1-939133-23-6},
pages = {397--411},
url = {https://www.usenix.org/conference/atc21/presentation/romero},
publisher = {USENIX Association},
month = jul
}

@inproceedings {serverlessinthewild-atc-2020,
author = {Mohammad Shahrad and others},
title = {Serverless in the Wild: Characterizing and Optimizing the Serverless Workload at a Large Cloud Provider},
booktitle = {2020 USENIX Annual Technical Conference (USENIX ATC 20)},
year = {2020},
isbn = {978-1-939133-14-4},
pages = {205--218},
url = {https://www.usenix.org/conference/atc20/presentation/shahrad},
publisher = {USENIX Association},
month = jul
}

@inproceedings{loongserve-sosp-2024,
author = {Wu, Bingyang and others},
title = {LoongServe: Efficiently Serving Long-Context Large Language Models with Elastic Sequence Parallelism},
year = {2024},
isbn = {9798400712517},
publisher = {Association for Computing Machinery},
address = {New York, NY, USA},
url = {https://doi.org/10.1145/3694715.3695948},
doi = {10.1145/3694715.3695948},
booktitle = {Proceedings of the ACM SIGOPS 30th Symposium on Operating Systems Principles},
pages = {640–654},
numpages = {15},
location = {Austin, TX, USA},
series = {SOSP '24}
}

@inproceedings {aplaserve-osdi-2023,
	author = {Zhuohan Li and others},
	title = {{AlpaServe}: Statistical Multiplexing with Model Parallelism for Deep Learning Serving},
	booktitle = {17th USENIX Symposium on Operating Systems Design and Implementation (OSDI 23)},
	year = {2023},
	isbn = {978-1-939133-34-2},
	address = {Boston, MA},
	pages = {663--679},
	url = {https://www.usenix.org/conference/osdi23/presentation/li-zhouhan},
	publisher = {USENIX Association},
	month = jul
}

@inproceedings{spotserve-asplos-2024,
author = {Miao, Xupeng and others},
title = {SpotServe: Serving Generative Large Language Models on Preemptible Instances},
year = {2024},
isbn = {9798400703850},
publisher = {Association for Computing Machinery},
address = {New York, NY, USA},
url = {https://doi.org/10.1145/3620665.3640411},
doi = {10.1145/3620665.3640411},
pages = {1112–1127},
numpages = {16},
location = {La Jolla, CA, USA},
series = {ASPLOS '24}
}

@inproceedings{exegpt-asplos-2024,
author = {Oh, Hyungjun and others},
title = {ExeGPT: Constraint-Aware Resource Scheduling for LLM Inference},
year = {2024},
isbn = {9798400703850},
publisher = {Association for Computing Machinery},
address = {New York, NY, USA},
url = {https://doi.org/10.1145/3620665.3640383},
doi = {10.1145/3620665.3640383},
pages = {369–384},
numpages = {16},
location = {La Jolla, CA, USA},
series = {ASPLOS '24}
}

@misc{pdserve-arxiv-2024,
      title={P/D-Serve: Serving Disaggregated Large Language Model at Scale}, 
      author={Yibo Jin and others},
      year={2024},
      eprint={2408.08147},
      archivePrefix={arXiv},
      primaryClass={cs.DC},
      url={https://arxiv.org/abs/2408.08147}, 
}

@misc{dejavu-arxiv-2024,
      title={D\'ej\`aVu: KV-cache Streaming for Fast, Fault-tolerant Generative LLM Serving},
      author={Foteini Strati and others},
      year={2024},
      eprint={2403.01876},
      archivePrefix={arXiv},
      primaryClass={cs.DC}
}

@misc{memserve-arxiv-2024,
      title={MemServe: Context Caching for Disaggregated LLM Serving with Elastic Memory Pool}, 
      author={Cunchen Hu and others},
      year={2024},
      eprint={2406.17565},
      archivePrefix={arXiv},
      primaryClass={cs.DC},
      url={https://arxiv.org/abs/2406.17565}, 
}

@inproceedings{grandslam-eurosys-2019,
author = {Kannan, Ram Srivatsa and others},
title = {GrandSLAm: Guaranteeing SLAs for Jobs in Microservices Execution Frameworks},
year = {2019},
isbn = {9781450362818},
publisher = {ACM},
address = {New York, NY, USA},
url = {https://doi.org/10.1145/3302424.3303958},
doi = {10.1145/3302424.3303958},
articleno = {34},
numpages = {16},
location = {Dresden, Germany},
series = {EuroSys '19}
}

@inproceedings {llumnix-osdi-2024,
author = {Biao Sun and others},
title = {Llumnix: Dynamic Scheduling for Large Language Model Serving},
booktitle = {18th USENIX Symposium on Operating Systems Design and Implementation (OSDI 24)},
year = {2024},
isbn = {978-1-939133-40-3},
address = {Santa Clara, CA},
pages = {173--191},
url = {https://www.usenix.org/conference/osdi24/presentation/sun-biao},
publisher = {USENIX Association},
month = jul
}



\end{document}